\documentstyle[12pt]{article}
\title{\huge Effect of the inhomogeneity of substrate on wetting transitions}
\author{\bf
 H. Ez-Zahraouy$^{*}$, L. Bahmad, and A. Benyoussef
\\
 Laboratoire de Magn\'{e}tisme et de la Physique
 des Hautes Energies
\\
Universit\'{e} Mohammed V, Facult\'{e} des Sciences, Avenue Ibn Batouta,\\
Rabat  B.P. 1014, Morocco
}
\date{ }
\begin{document}
\maketitle

\begin{abstract}
\mbox{~~~  }
We study the substrate inhomogeneity effect on the wetting of a
spin-$1/2$ Ising ferromagnetic film in an external magnetic field $H$,
using Monte Carlo simulations. It is found that the inhomogeneity leads to
the formation of clusters of positive and negative spins in each layer. However, depending on the values of $H$, for a fixed surface
magnetic field $H_s$, each layer exhibits three different phases;
totally wet, no wet and partially wet.
In the latter case, we show the existence of three distinct configurations
namely: A configuration  in which the layer is partially wet with a
total disconnection between clusters (PWTD);
 a configuration in which the layer is partially wet with a partial
disconnection between  clusters (PWPD);
the  configuration in which the layer is partially wet and the clusters
 are totally connected (PWTC).
Furthermore, we show that an increase of $H_s$ values breaks the bonds
connecting some clusters of the phase (PWTC) which leads to an expansion of
the (PWPD) region.
On the other hand, the frequency distribution and the average size
of clusters are investigated in the (PWTD) region for fixed values of temperature $T$, $H_s$ and  $H$. Moreover, we show the existence of $3D$-islands resulting from the formation of islands in each layer.

\end{abstract}
----------------------------------- \\
{\it Keywords:}  Wetting; Monte Carlo simulations; Inhomogeneity;
Layering; island distribution. \\
\mbox{~}(*) Corresponding author: ezahamid@fsr.ac.ma \\

%\newpage

\section{Introduction}
\mbox{  }

Multilayer films adsorbed on attractive substrates may exhibit a variety of
possible phase transitions, as has been reviewed by Pandit {\it et al.} [1],
Pandit and Wortis [2], Ebner {\it et al.} [3] and Patrykiejew {\it et al.}
[4]. The existence of phase diagrams depending on the strength of the substrate
potential in a lattice gas model for multilayer adsorption, have been shown by
Benyoussef and Ez-Zahraouy [5].
The wetting and layering transitions are caused by the competition between
short-ranged attractions and long-ranged repulsions in systems like langmuir
monolayers and adsorbed monolayers [6].
These transitions have also been found, under the effect of the geometry of
the system, in the case of the double wedge wetting [7] in magnetic films,
and the wetting behavior of associating binary mixtures at attractive walls [8].
We showed recently, in Ref. [9], the existence of these transitions under the
effect of the surface coupling, when using the mean field theory. \\
On the other hand, many recent studies showed that indirect interactions can
lead to the formation of nano-structures at surfaces [10] mediated by the
structure and the geometry of the film.
Monte Carlo Simulations [11] and subsequently derived mean field theories
[12], showed the existence of higher island densities than those expected
by standard nucleation theory.
It is shown that several experimental works, e.g. [13], supports these
predictions, such as the deposition of $Au$ on mica substrates at a high
temperature [14]. \\
A small variation of the preparation conditions may considerably change
the obtained magnetic structure and allows the investigation of atomic
structure as well as the magnetic domain structure of several nanostructured
systems [15,16].
However, Lenz and Lipowski [17] showed that the morphology of wetting
layers on structured surfaces is determined by the geometry of the underlying
surface domains, and that the wetting layer exhibits several distinct
morphologies. \\
\underline{On the other hand, it has been shown [18] in the case of a two-dimensional lattice-gas model,} \\
\underline{ that some basic regimes of drying mediated nanoparticle assembly are distinguished by the } \\
\underline{ spatial uniformity of solvent dynamics, and by the fluctuations of nanoparticle domain}\\
\underline{ boundaries following evaporation. It has also shown that when solvent disappears}\\
\underline{ homogeneously from the surface, disk-like or ribbon- like domains reminiscent of spinoidal} \\
\underline{ decomposition form at early time. If instead domain boundaries are frozen following} \\ 
\underline{ evaporation, dynamical constraints arrest this growth at an early stage. Network structures} \\
\underline{ are formed as vapour nuclei meet [18], otherwise networks fragment to form distinct}\\
\underline{ domain that asymmetrically evolve as in homogeneous coarsening. }
 \\
Our aim in this work is to study the  effect of the inhomogeneity of
substrate on the wetting transitions of a spin-$1/2$ Ising film when a
surface magnetic field is applied on alternate clusters of the surface.
The paper is organized as follows. In section $2$, we describe the model
and the method used: Monte Carlo (MC) simulations. In section $3$ we present
results and discussions. \\

\section{Model and Monte Carlo simulations}
\underline{Our model can describe either a magnetic system in which the Ising spin variables are } \\
\underline{$S_i=+1,-1$, or a physical state of the liquid-gas model where the occupation variables} \\
\underline{ $ n_i=+1,0$ (+1 when the site is occupied and 0 if the site is not occupied) can replace }\\
\underline{the spin variables when $n_i=(1+S_i)/2$. The external magnetic field $H$ and the surface }\\
\underline{magnetic field $H_si$ can then play  the role of the chemical potential and the substrate potential,}\\
\underline{ respectively. These analogies concerning the lattice gaz model and a magnetic system are more }\\
\underline{explained in Refs. [1-4]. The case a two-dimensional lattice gas model is studied in details in [18].}\\
The system we are studying, illustrated by Fig. 1, is a magnetic thin film
formed with $N=4$ layers coupled ferromagnetically.
Each layer is a square of dimension $N_x \times N_y=64 \times 64$ spins.
$N_x$ and $N_y$ stand for the number of spins in the $x$ and $y$ directions,
respectively.
A surface magnetic field $H_s$ is acting only on alternate clusters of spins
of the surface $k=1$.
This is represented, in Fig. 1, by symbols $(+)$ for alternate clusters of
dimension $l_p \times l_p=8 \times 8$. Whereas $H_s$ is absent for the remaining
alternate clusters of dimension $l_n \times l_n=8 \times 8$. These clusters
are represented by symbols $(o)$. \\
The Hamiltonian governing this system is given by
\begin{equation}
{\cal H}=-\sum_{<i,j>}J_{ij}S_{i}S_{j}-\sum_{i}(H+H_{s_i})S_{i}
\end{equation}
where, $ S_{l}(l=i,j)=-1,+1$ are the random spin variables and \\
\underline{the interaction $J_{ij}$ between different spins is assumed to be constant $J_{ij}=J$. } \\
\underline{The first summation in Eq. (1) includes only the nearest neighbor spins. The second } \\ 
\underline{summation runs over all sites $i$ of the system. $H$ is an external magnetic field applied}\\ 
\underline{ on each site of the film. }\\
The surface magnetic field $Hs_i$, 
\underline{appearing in the Hamiltonian $(1)$,}
 is applied only on sites $i$ of the surface $k=1$, and distributed alternatively, \underline{according to}:
\begin{equation}
H_{s_i}=\left\{
	\begin{array}{ccc}
	+H_s &  \mbox{for all sites i $\epsilon$} & \mbox{clusters with
	symbols (+) } \\
	0    &  \mbox{for all sites i $\epsilon$} & \mbox{clusters with
	symbols (o) }.
	\end{array}
    \right.
\end{equation}
\underline{The inhomogeneity in the surface results from this distribution of the magnetic field $Hs_i$.} \\
\underline{We use the Monte Carlo simulations under the Metropolis algorithm. }\\
\underline{An attempt to flip a randomly chosen spin in site $'i'$ from up to down, or vice versa,}\\
\underline{is envisaged using the Metropolis probability:}
\begin{equation}
p_M=min[1,exp(-\Delta E/k_B T)],
\end{equation}
\underline{where, $\Delta E$ is the resulting change in energy after the flip of the spin $'i'$, $k_B$ is }\\
\underline{the Boltzmann constant and $T$ the absolute temperature. The flip of the spin is accepted}\\
\underline{ if $p_M$ is greater than a random number $p_r$ generated by the computer so that $p_r \epsilon [0,1]$.}\\
\underline{ Otherwise the flip of the spin located at the site 'i' is ignored if the randomly generated}\\
\underline{ number is so that $p_r > p_M$. One Monte Carlo step (MCS) is reached once the number}\\
\underline{ of visited sites is close to the number of sites of the system. For a fixed temperature and/or}\\
\underline{ surface magnetic field, runs of $5\times 10^6$ Monte Carlo steps (MCS) are performed for }\\
\underline{ different initial conditions. Mean values of computed quantities are estimated for different}\\
\underline{ initial conditions, with the discard of the first $5\times 10^5$ (MCS) for each run.}  \\
A preliminary study showed that the topologies of the phase diagrams are
similar (with different numerical values) when the film thickness varies from $N=3$
\underline{to $N=9$ layers}, when the number of spins of each layer varies from
$N_x=N_y=16$ to $N_x=N_y=64$; where $N_x$ and $N_y$ are the number, of spins of
each layer, in the $x$ and $y-$directions, respectively.  
\underline{When increasing the film thickness, the topology of the established phase diagrams are }\\
\underline{similar for thin films $(N<10)$ layers. For simplicity and fast Monte Carlo simulations, and }\\ 
taking into account the above considerations we give numerical results in this paper for a film with $N=4$ layers and $N_x=N_y=64$ spins for each layer.
\\
We mean by an 'island', a set of negative spins surrounded by positive ones.
In all the following, our interest will be concentrated on the mean size of
islands (MSI) and the average number of islands (ANI).
The different phases are determined by values of the average number of islands
and their mean sizes as following: \\
* The PWTD phase is characterized by:
$ ANI \le 32 $ and $ MSI \le (MSI \le N_x \times N_y)/2 $. \\
* The PWPD phase is present for:
$ANI \le 32$ and $  (N_x \times N_y)/2 \le  MSI \le N_x \times N_y $. \\
* The PWTC phase corresponds to:
$ANI \ge 1$ and $ MSI \ge (N_x \times N_y)/2 $. \\
However, in the particular case of the non wet phase, we have $ ANI=1 $ and
$MSI= Nx \times Ny$  (which is the total number of spins of each layer).
We note that in this particular case , although positive spins are
completely absent in the layer, we assume that this situation corresponds
to a unique island.

\section{Results and discussion}
\mbox{~~~}
A sketch of the geometry of the system we are studying, is presented
in Fig. 1. In order to outline the behaviour of the 
\underline{mean size of islands (MSI) and the average number of islands (ANI)}
as a function of the external magnetic field, we plot in Figs. 2a and 2b, these parameters, for
a surface magnetic field $H_s=2.5$ and a temperature $T=3.5$.
Indeed, Fig. 2a corresponds to the surface $k=1$, whereas Fig. 2b is plotted
for the last layer $k=N=4$.
It is found that, for the surface $k=1$ (Fig. 2a) and the last layer
$k=N$ (Fig. 2b), when
decreasing the external magnetic field from $H=0$, 
\underline{the average number of islands (ANI)} exhibits two
islands for $H \approx -0.1$, with a rapid increasing of the number of
negative spins in each layer, as it is illustrated by 
\underline{the mean size of islands (MSI)} curve.
For $T=3.5$ and $-0.16 < H < -0.07$, the \underline{average number of islands (ANI)} increases rapidly and the
\underline{mean size of islands (MSI)} undergoes a local maximum.
When decreasing the magnetic field $H$ so that $H < -0.20$
the average number of islands of the surface stabilizes at $32$, whereas it drops suddenly to
one cluster formed with negative spins, in the case of the last layer $k=N$.
In the case of the surface, the \underline{mean size of islands (MSI)} undergoes a large
local plateau corresponding to the value $64$, followed by the apparition of
a unique island for $H \approx -0.80$.
This scenario is absent in the case of the last layer $k=N$, Fig. 2b, where the
islands are rapidly connected to each other for $H < -0.15$.
On the other hand, the increasing surface magnetic field effect, on island
formation is illustrated by Figs. 3a and 3b for a fixed
temperature $T=3.5$ for several values of the surface magnetic field:
$H_s=1.5$, $H_s=2.0$ and $H_s=2.5$.
Indeed, for small values of the external magnetic field, $-0.20 < H < 0$, the increasing
surface magnetic field $H_s$ is not felt at this "higher" temperature value $3.5$.
While, for $H < -0.20$, the increasing surface magnetic field amplitude is
to produce a large step of mean size island value $64$, see Fig. 3a.
It is worth to note that this step disappears for very small surface magnetic
field values.
Concerning the average number islands (ANI), the same
arguments still valid when increasing the surface magnetic field amplitude
$H_s$, Fig. 3b. The islands are rapidly connected to each other for very
small values of $H_s$.
On the other hand, we show in Fig. 4 the existence of three regions of
island configurations in the plane $(T,H)$, for $k=N$ and a fixed surface
magnetic field $H_s=2.0$.
These three different regions are: the totally wet (TW) region, the no wet
region (NW) and the partially wet region.
In the latter case, see in Fig. 4, the existence of three distinct
island configurations, namely; the (PWTD) configurations
where the layer is partially wet with a
total disconnection of the islands; the (PWPD) configurations in which the
layer is partially wet with a partial disconnection of the islands;
The (PWTC) configurations corresponding to a partially wet layer with
totally connected clusters.
Indeed, at low temperature values, only the totally wet
(TW) phase for $H \approx 0$, and the non wet (NW) phase for $H \le -0.5$
are present.
At higher temperature values, provided that $T$ is kept less than the
critical value $T_c=3.87$, the partially wet phase occurs.
Starting from the totally wet phase (TW), present for positive values of $H$,
the decreasing of the external magnetic field (from $0$ to negative values),
gives rise to the phase (PWTD) showing totally disconnected configuration of
islands. This phase is followed by the (PWPD) phase in which the configuration
of islands are partially disconnected when $-0.16 < H < -0.07$.
For $H \le -0.20$, the partially wet phase (PWTC) where the islands are totally
connected to each other, is reached.
Decreasing the magnetic field $H$ more and more, the non wet phase
$(NW)$ is found and persists for $H \rightarrow -\infty$. 
\underline{The topology of the established phase diagrams stand valid for the other layers $k<N$}\\
\underline{ of the film. The layer k=4 is chosen because the studied phases in the partially wetting region:}\\
\underline{ totally disconnected (PWTD), partially disconnected (PWPD), and totally connected (PWTC)}\\
\underline{ are well outlined in this layer. Moreover this layer is chosen in order to outline the range effect}\\
\underline{ of the substrate on deeper layers.} \\
In order to more clarify these findings, we plot in Figs. 5a and 5b, different
island configurations in different regions of the phase diagram for the different layers of the film. \\
Indeed, Fig. 5a, shows that, for $H_s=2.0$, $T=3.5$ and $H=-0.13$, the first layer $k=1$ 
\underline{shows a configuration corresponding to a partially wet layer with totally connected clusters}\\ (PWTC). 
The second layer $k=2$ belongs to \underline{the partially disconnected configuration} (PWPD), whereas  the third layer
$k=3$ is in the \underline{totally disconnected phase} (PWTD). 
On the other hand, when maintaining the parameters $T$ and $H_s$ constant, and varying only
the external magnetic field, a given layer undergoes different configurations
belonging to different regions: totally connected (PWTC), partially disconnected (PWPD) and totally disconnected (PWTD). This situation is illustrated by Fig. 5b for the last layer $k=4$. Indeed, for $H=-0.13$ this
layer is in the \underline{totally connected phase} (PWTC). For $H=-0.15$ the configuration of this layer \underline{partially disconnected} (PWPD); and when $H=-0.17$ the islands of this layers are completely disconnected from each other so that this configuration belongs to \underline{totally disconnected} (PWTD) region.
The numerical values given in this work correspond to $N_x=N_y=64$. However,
a preliminary study showed that when $N_x=N_y=32$ and $T=3.5$ the
partially disconnected region (PWPD) is located in the interval $-0.13 < H < -0.09$. Hence, the finite
size effect and thermodynamic limit is to enlarge partially wetting regions (PW).
Moreover, the formation of islands in each layer leads to the formation of
the three dimensional islands ($3D$-islands). The superposition of these $2D$-island
of adjacent layers leads to the formation of such $3D$-islands in the film.
For example, Fig.5a shows these  $3D$-islands, when superposing the layers $k=1,2,3$
over each other. This situation is more outlined, for example, in the rectangle of
coordinates in the rectangle $16 \le x \le 24$ and $32 \le y \le 40$
for the layers $k=1,2,3$ and $4$, respectively.
The superposition of these islands constitute a three dimensional island in
the bulk of the film.
However, the height of the formed islands depends on the values of
temperature, surface and  external magnetic fields.
In the example of Fig. 5a, some island heights are close to 4.
A preliminary study showed that the island height decreases with decreasing
values of $H_s$, for fixed values of temperature $T$ and external magnetic
field $H$.
To complete this study, the frequency distribution of different configurations,
are plotted in Fig. 6. This figure shows the corresponding frequency island size distributions,
for the totally disconnected region (PWTD), for the layers $k=1$, $k=2$, $k=3$ and $k=4$ when the 
temperature is fixed at $T=3.5$, the surface magnetic field is $H_s=2.0$ and $H=-0.142$.
The different scenarios presented by the system concerning
the inhomogeneity substrate effect on the distribution
of cluster sizes and number of existing clusters are presented. Indeed, the
mean size cluster of negative spins undergoes
a local maximum before exhibiting  first and second steps. The plateau of
the first step increases with increasing surface magnetic field values. \\

\section{Conclusion}
We have studied the effect of the inhomogeneity of a substrate on the
wetting transitions of a spin-$1/2$  ferromagnetic Ising thin film
under the effect of an alternate surface magnetic field $H_s$ acting on
alternate clusters of the
surface, using Monte Carlo simulations. In the partial wetting region each
layer exhibits three different configurations namely:
partial wetting with totally disconnected cluster (PWTD), partial wetting with
partially disconnected clusters (PWPD) and partial wetting with totally
connected clusters (PWTC).
On the other hand, we found that increasing the surface magnetic field
leads to the disconnection of some connected clusters in the  (PWPD) region.
Furthermore the $3D$-islands occur for strong surface magnetic field values.
Moreover, the distributions of the cluster size, in each layer of the film,
are also computed. \\
\\
%\newpage
\noindent{ \Large \bf References}
\begin{enumerate}

\item[{[1]}] R. Pandit, M. Schick and M. Wortis, Phys. Rev. B {26} (1982) 8115.

\item[{[2]}] R. Pandit and M. Wortis, Phys. Rev. B { 25}  (1982) 3226.

\item[{[3]}] C. Ebner and W. F. Saam, Phys. Rev. Lett. { 58} (1987) 587.

\item[{[4]}] A. Patrykiejew A., D. P. Landau and K. Binder, Surf. Sci. { 238}  (1990) 317.

\item[{[5]}] A, Benyoussef, H. Ez-Zahraouy, Physica A { 206}, 196(1994);
ibid, J. Phys. I France 4  (1994) 393.

\item[{[6]}] A. D. Stoycheva, and S. J. Singer, Phys. Rev. Lett. { 84} (2000)  4657.

\item[{[7]}] L. Bahmad, A, Benyoussef, and H. Ez-Zahraouy, Phys. Rev. E
 { 66}  056117 (2002); A. Milchev, M. Muller, and D. P. Landau,
 Phys. Rev. Lett. { 90}  (2003) 136101-1.

\item[{[8]}] A. Patrykiejew, L. Salamacha, S. Sokolowski, and O. Pizio,
Phys. Rev. E { 67}  (2003) 061603.

\item[{[9]}] L. Bahmad, A. Benyoussef and H. Ez-Zahraouy,
Surf. Sci. { 536}  (2003) 114.

\item[{[10]}] K. A. Fichthorn and M. L. Merrick, Phys. Rev. B { 68}
 (2003) 041404 (R).

\item[{[11]}] S. Ovesson, A. Bogicevic, G. Wahnstrom, and B. I. Lundqvist,
Phys. Rev. B { 64}  (2001) 115423.

\item[{[12]}] J. A. Venables and W. Kohn Phys. Rev. B { 66}  (2002) 195404. 

\item[{[13]}] H. Brune, K. Bromann, K. Kern, J. Jacobsen, P. Stolze,
K. Jacobsen, and J. Norskov, Phys. Rev. B { 52}  (1995) R14 380.

\item[{[14]}] K. Reichelt and H. O. Lutz, J. Cryst. Growth  { 10}
(1971) 103.

\item[{[15]}] R. Allenspach, M. Stampanoni, and A. Bischof, Phys. Rev. Lett.
{\bf 65}  (1990) 3344.

\item[{[16]}] A. Kubetzka, O. Pietzsch, M. Bode, and R. Wiesendanger,
Phys. Rev. B { 63} (2001) 140407(R).

\item[{[17]}] P. Lenz and R. Lipowski, Phys. Rev. Lett. { 80}  (1998) 1920.

\item[{[18]}] \underline{E. Rabani, D. R. Reichman, P.L. Geissier and L. E. Brus, Nature { 426} (2003) 271.}

\end{enumerate}

%\newpage
\noindent{\bf Figure Captions}\\

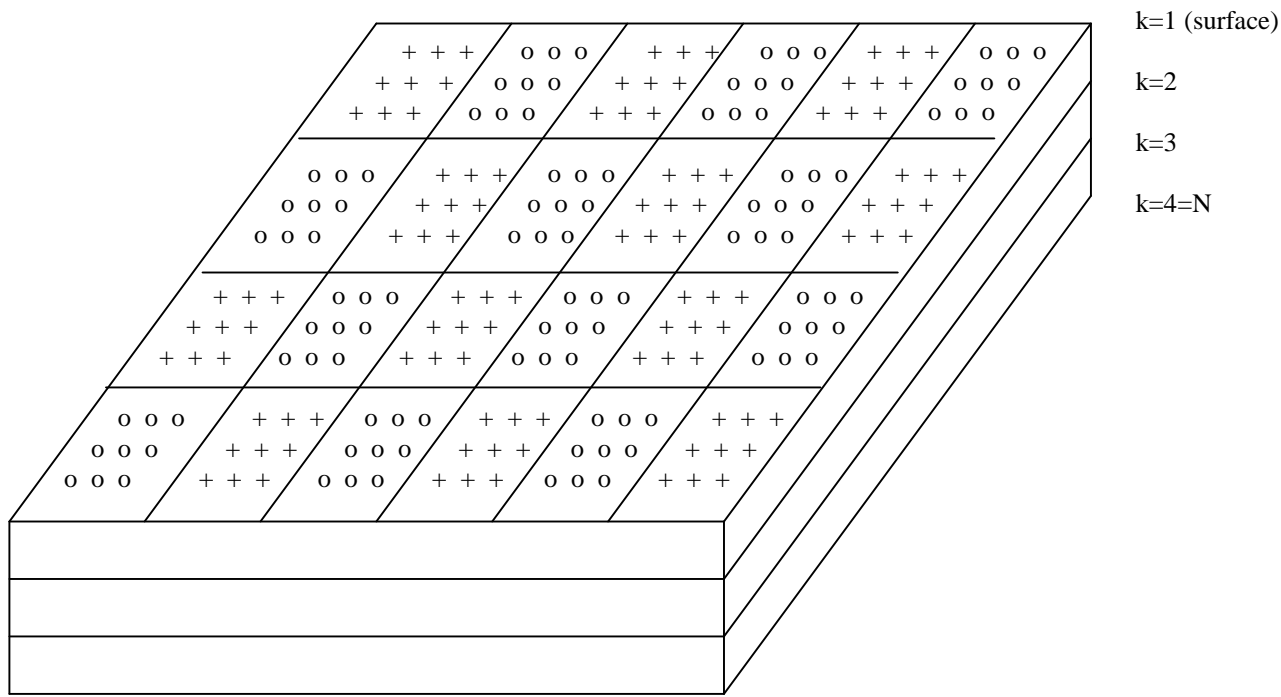
\noindent{\bf Figure 1.}: \\
A geometry sketch of the studied system formed with $N=4$ layers.
A surface magnetic field $H_s$ is acting on alternate cluster spins of
the surface $k=1$. $H_s$ is present on islands with symbols: $'+'$, and absent
elsewhere: islands with symbols $'o'$. \\

\noindent{\bf Figure 2}: \\
The dependency of the mean size of islands (MSI), and the average
number of islands (ANI) of the surface $k=1$ (a), and the last layer $k=N$,
as a function of the external magnetic field $H$ for a fixed temperature
$T=3.5$ and a surface magnetic field value $H_s=2.5$. \\

\noindent{\bf Figure 3}:\\
(a) The mean size of islands (MSI), and (b) the average number of
islands (ANI) of the surface $k=1$ as a function of $H$ for $T=3.5$ and
several values of $H_s$: $1.5$; $2.0$ $2.5$. \\

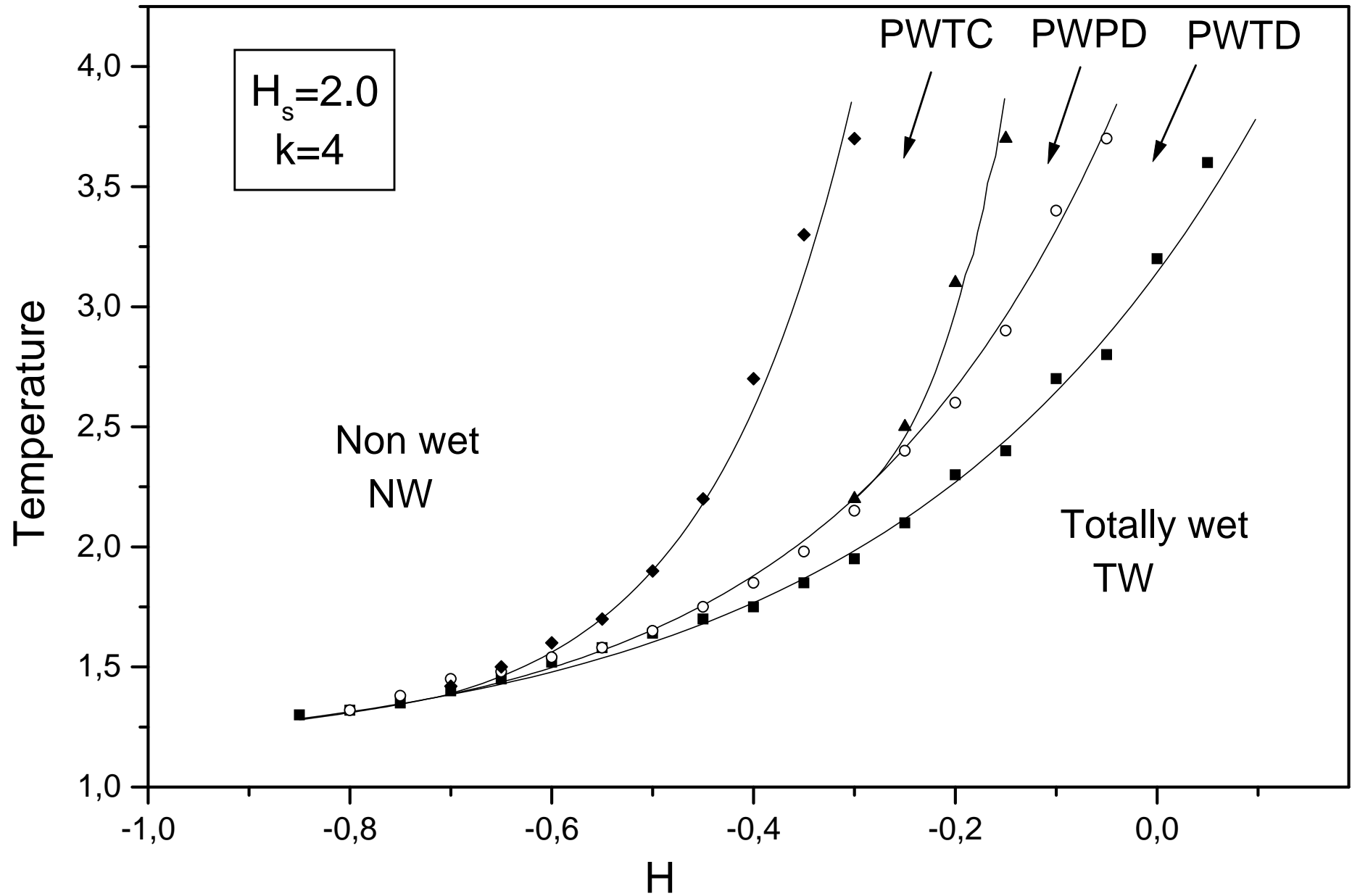
\noindent{\bf Figure 4}: \\
The ($T,H$) phase diagram for the layer $k=4$, as a
function of the external magnetic field $H$, for several values of
the surface magnetic field $H_s$: $1.5$; $2.0$ and $2.5$. The regions
corresponding to the different configurations: (TW) totally wet, (PWTC)
partially wet with totally connected clusters, (PWPD)
partially wet with partially disconnected clusters, (PWTD)
partially wet with totally disconnected clusters, and (NW) non wet,
are presented. \\

\noindent{\bf Figure 5}: \\
island maps of positive spins for different layers of the film with $N=4$
a) The first layer $k=1$ is in the (PWTC) configuration,
the second layer $k=2$ is in the configuration (PWPD), while the layer
$k=3$ belongs to the configuration (PWTD); for  $T=3.5$, $H=-0.13$ and $H_s=2.0$. \\
b)For $H=-0.13$, $H=-0.15$
and $H=-0.17$, the last layer $k=4$ exhibits  three different configurations
(PWTC), (PWTD) and (PWTD), respectively, for  $T=3.7$ and $H_s=2.0$. \\

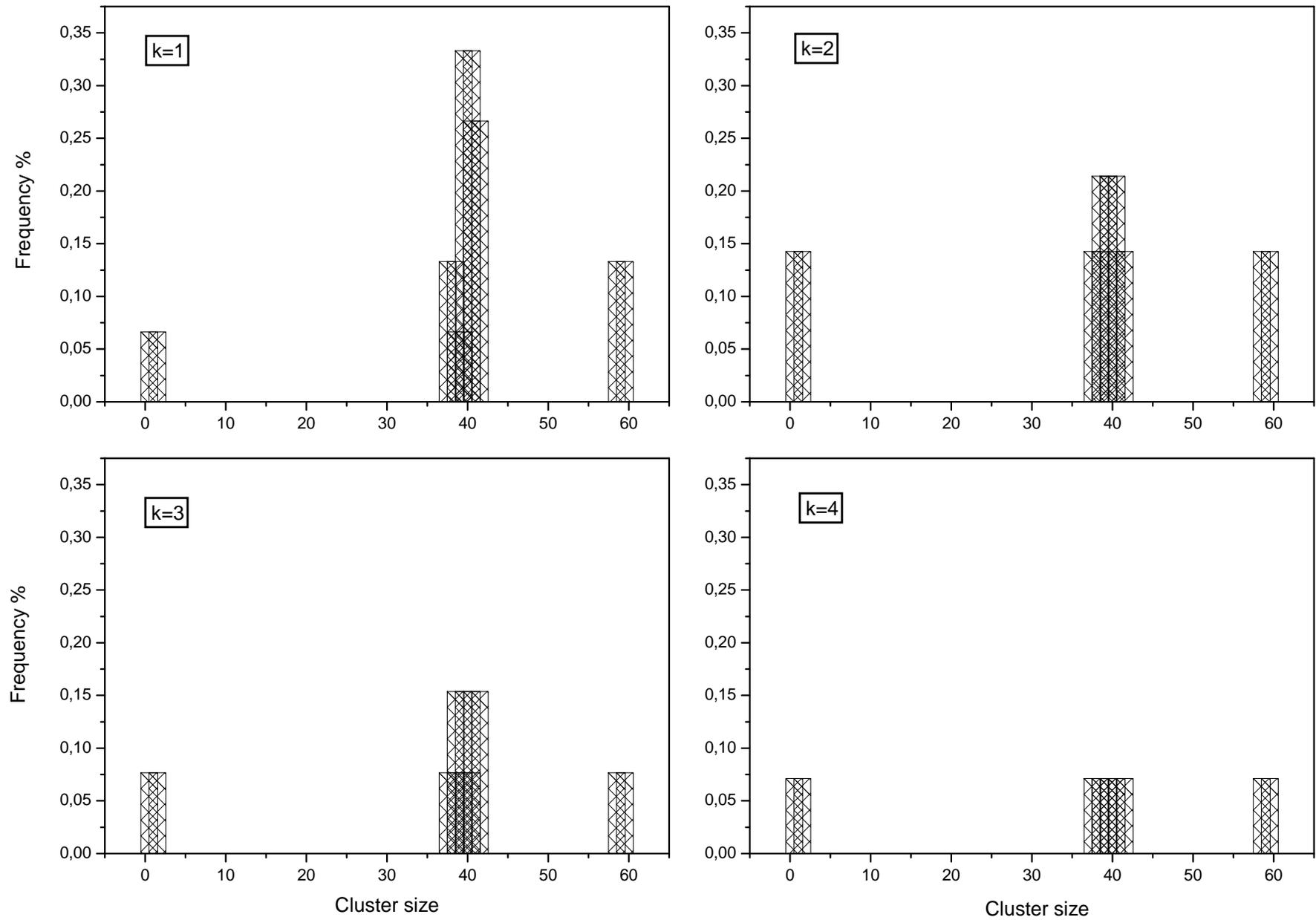
\noindent{\bf Figure 6}: \\
Frequency island size distributions for the layers $k=1$, $k=2$, $k=3$ and
$k=4$ for the configuration (PWTD) corresponding to $T=3.5$,
$H_s=2.0$ and  $H=-0.142$.

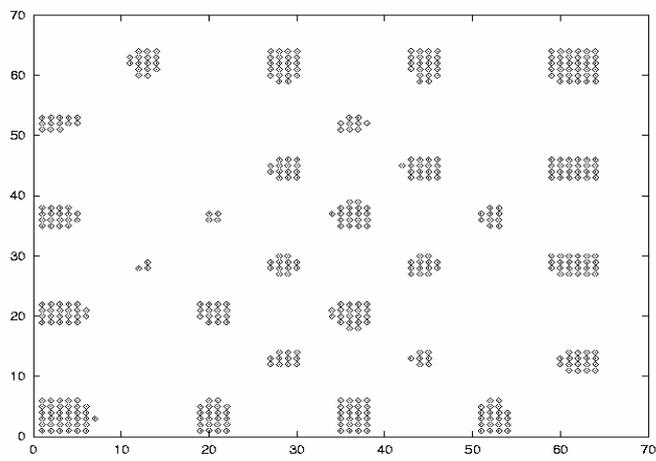
\end{document}